\documentclass{emulateapj}

\begin{document}

\shortauthors{Luhman et al.}
\shorttitle{Spitzer Survey of Taurus}

\title{A Survey for New Members of Taurus with the Spitzer Space Telescope}

\author{
K. L. Luhman\altaffilmark{1},
B. A. Whitney\altaffilmark{2},
M. R. Meade\altaffilmark{3},
B. L. Babler\altaffilmark{3},
R. Indebetouw\altaffilmark{4},
S. Bracker\altaffilmark{3},
and E. B. Churchwell\altaffilmark{3}}

\altaffiltext{1}{Department of Astronomy and Astrophysics,
The Pennsylvania State University, University Park, PA 16802;
kluhman@astro.psu.edu.}

\altaffiltext{2}{Space Science Institute, 4750 Walnut Street, Suite 205,
Boulder, CO 80301.}

\altaffiltext{3}{Department of Astronomy, The University of Wisconsin, 
475 North Charter Street, Madison, WI 53706.}

\altaffiltext{4}{Department of Astronomy, University of Virginia, 
P.O. Box 3818, Charlottesville, VA 22903-0818.}

\begin{abstract}

We present the results of a search for new members of the Taurus star-forming
region using the Infrared Array Camera (IRAC) aboard the {\it Spitzer 
Space Telescope}. With IRAC images of 29.7~deg$^2$ of Taurus at 
3.6, 4.5, 5.8, and 8.0~\micron, we have
identified sources with red mid-infrared colors indicative of disk-bearing
objects and have obtained optical and infrared spectra of 23 of these 
candidate members. 
Through this work, we have discovered 13 new members of Taurus, two of which 
have spectral types later than M6 and thus are likely to be brown dwarfs 
according to the theoretical evolutionary models of Chabrier and Baraffe.
This survey indicates that the previous census of Taurus has a completeness
of $\sim80$\% for members with disks.
The new members that we have found do not significantly modify the previously
measured distributions of Taurus members as a function of position, mass, 
and extinction. For instance, we find no evidence for a population of 
highly reddened brown dwarfs ($A_K\sim2$) that has been missed by previous
optical and near-infrared surveys, which suggests that brown dwarf disks
are not significantly more flared than disks around stars. 
In addition to the new members, we also present IRAC 
photometry for the 149 previously known members that appear within this survey,
which includes 27 objects later than M6.

\end{abstract}


\section{Introduction}
\label{sec:intro}

As one of the nearest sites of active star formation ($d=140$~pc), the Taurus
complex of dark clouds has long been a popular laboratory for 
studies of the birth of stars and planets.
A fundamental prerequisite for most observations of Taurus is a 
census of its stellar content. 
Over the decades, a variety of methods have been employed to identify 
members of Taurus, including 
objective prism spectroscopy of H$\alpha$ \citep{joy46,joy49,bri93,bri99},
X-ray imaging with the {\it Einstein Observatory} \citep{fei87,wal88}
and the {\it R\"ontgen Satellite} \citep{ss94,neu95,wic96,bri99},
proper motion measurements \citep{jh79,har91,gom92},
near- and mid-infrared (IR) photometry from
ground-based telescopes \citep{gom94,itg96,lr98} and 
the {\it Infrared Astronomical Satellite (IRAS)} \citep{bei86,ken90,ken94},
and wide-field optical and near-IR imaging
\citep{bri98,bri02,luh00tau,luh04tau,luh06tau,luh03tau,mar01,gui06}.
The census of known members of Taurus has steadily grown \citep{ck79,hb88,kh95}
and now contains more than 300 stars and brown dwarfs.

The {\it Spitzer Space Telescope} \citep{wer04}
offers a combination of sensitivity and large
field of view that is unprecedented for a mid-IR telescope, making it
uniquely suited for finding disk-bearing objects down to low masses 
($M\sim0.01$~$M_\odot$), through high levels of extinction ($A_V\sim100$),
and across large areas of sky ($\theta\sim10\arcdeg$).
To take advantage of this capability, we have used archival
{\it Spitzer} images encompassing 29.7~deg$^2$ of Taurus to perform a search 
for new members of the region.
In this paper, we present the collection and analysis of the 
{\it Spitzer} data (\S~\ref{sec:irac}), the identification of 
candidate members with these data (\S~\ref{sec:cand}), 
the spectroscopy and classification of these candidates (\S~\ref{sec:spec}),
an evaluation of the completeness of this survey and previous ones
(\S~\ref{sec:complete}), and the implications of our updated census 
for the distribution of Taurus members as a function of position, mass,
and extinction and the implications of our mid-IR photometric catalog 
for the frequency of disks in Taurus (\S~\ref{sec:imp}).

\section{Analysis of Spitzer Data}
\label{sec:irac}

\begin{figure}
\plotone{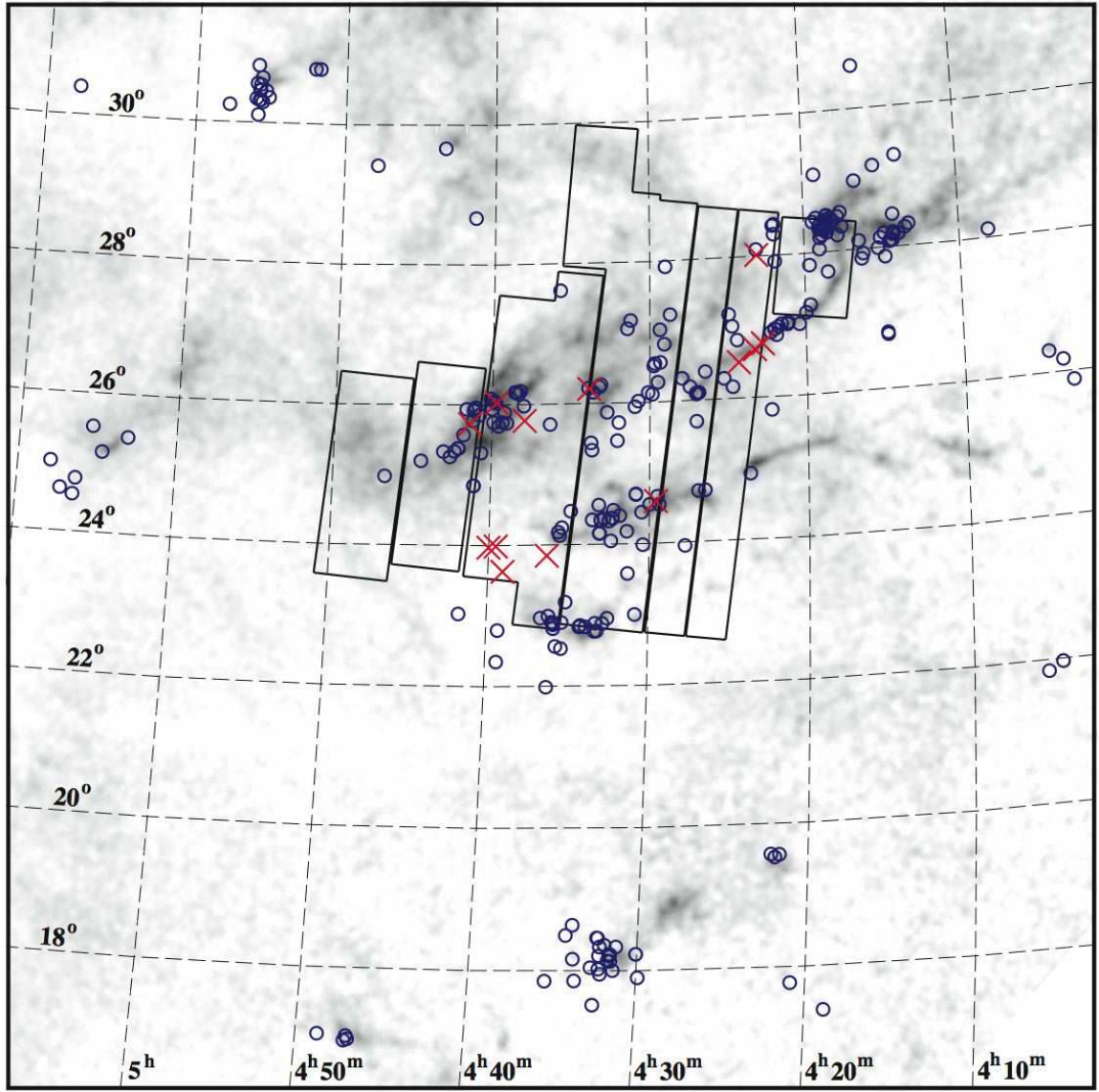}
\caption{
Spatial distribution of previously known members of the Taurus
star-forming region ({\it circles}) and new members discovered in this work 
({\it crosses}, Table~\ref{tab:new}) shown with a map of extinction 
\citep[{\it grayscale},][]{dob05}. The area imaged with IRAC aboard the 
{\it Spitzer Space Telescope} in this work is indicated ({\it solid lines}).}
\label{fig:map}
\end{figure}

To search for new members of Taurus, we used unpublished archival images 
at 3.6, 4.5, 5.8, and 8.0~\micron\ that were obtained with the Infrared Array 
Camera \citep[IRAC;][]{faz04} aboard the {\it Spitzer Space Telescope} 
during the General Observer program 3584 by D.\ Padgett. 
The plate scale and field of view of IRAC are $1\farcs2$ and
$5\farcm2\times5\farcm2$, respectively. The camera produces images with
FWHM$=1\farcs6$-$1\farcs9$ from 3.6 to 8.0~\micron.
The IRAC data for Taurus were collected between 2005 February 20 and 27 (UT).
The images were obtained in 16 adjacent maps, each of which  
contained a mosaic of pointings separated by $290\arcsec$ 
and aligned with the array axes.  At each cell in the map, images were 
obtained in the 12~s high dynamic range (HDR) mode, which provided one 0.4~s 
exposure and one 10.4~s exposure.  Each of the 16 maps was observed twice.
These observations encompassed a total area of 29.7~deg$^2$.
The boundaries of the imaged fields are indicated in the map of Taurus
in Figure~\ref{fig:map}.

The IRAC images were processed with the S11.4.0 pipeline from the {\it Spitzer}
Science Center and the pipeline for 
the Galactic Legacy Infrared Mid-Plane Survey Extraordinaire 
\citep[GLIMPSE,][]{ben03} after it was modified to enable the processing 
of HDR frames. Sources were extracted from individual frames 
using a modified version of DAOPHOT 
\citep{ste87}\footnote{For more details concerning the photometry routines
and GLIMPSE pipeline processing, see
the GLIMPSE documents at http://www.astro.wisc.edu/glimpse/docs.html.}.
Before the photometry was measured, image defects were corrected and saturated
stars were masked. The image defects included column pulldown and muxbleed
for 3.6 and 4.5~\micron\ and banding for 5.8 and 8.0~\micron\ \citep{hor04}.
We did not mask cosmic rays because overmasking of cosmic rays can modify 
stellar fluxes. An array-location-dependent photometric correction was applied 
to the IRAC photometry as specified by the {\it Spitzer} Science Center 
\footnote{http://ssc.spitzer.caltech.edu/irac/locationcolor}.
The sources were merged across observations and wavelengths 
using the {\it Spitzer} Science Center bandmerger, as modified by GLIMPSE.
At this step, the IRAC sources were merged with near-IR sources from the 
Point Source Catalog of the Two-Micron All-Sky Survey \citep[2MASS,][]{skr06}.
To prevent cosmic rays from being falsely included in our source lists, 
we required detections of each source in at least two adjacent IRAC bands
with a signal-to-noise ratio (SNR) greater than 5 in at least one and two 
bands for 0.4 and 10.4~sec, respectively.
These criteria were developed through analysis of color-color and 
color-magnitude diagrams and comparison of 0.4 to 10.4~sec data for
the Taurus data in this work and observations of the Large Magellanic Cloud
by \citet{mei06}.
Even if the source is reliable, the photometry in all bands may not be. 
Therefore, for both the 0.4 and 10.4~sec data, we rejected the flux in a band 
if SNR$\leq5$ in the combined measurement from the two exposures for that band.
For the 10.4~sec data, 
we excluded photometry brighter than 9.5, 9.0, 6.5, and 6.5~mag
at 3.6, 4.5, 5.8, and 8.0~\micron, respectively.
For the 0.4~sec data, we flag as upper limits 
the magnitudes brighter than 6.0, 5.5, 3.0, and 3.0~mag, respectively.
When merging the catalogs for the short and long exposures, we matched
a pair of sources if the separation was less than $0\farcs7$. For these
objects, we compared the short and long exposure data at a given band
and adopted the measurement with the smallest error. 
The list of IRAC sources, which we refer to as the Taurus IRAC Point
Source Archive (or ``the Archive"), contains $\sim450,000$ sources and will be
available to the public\footnote{http://www.astro.wisc.edu/glimpse/Taurus}.

We used the calibration from \citet{rea05}, which differs slightly from the 
calibration applied to previous IRAC measurements in Taurus by \citet{har05}.
After adjusting the data from \citet{har05} to the calibration from 
\citet{rea05} and comparing photometry for $\sim40$ sources appearing in both 
studies, we find that the average photometry for each band agrees between the
two studies to within $\sim2$\%.

\section{Selection of Candidate Members of Taurus}
\label{sec:cand}

\begin{figure}
\epsscale{.6}
\plotone{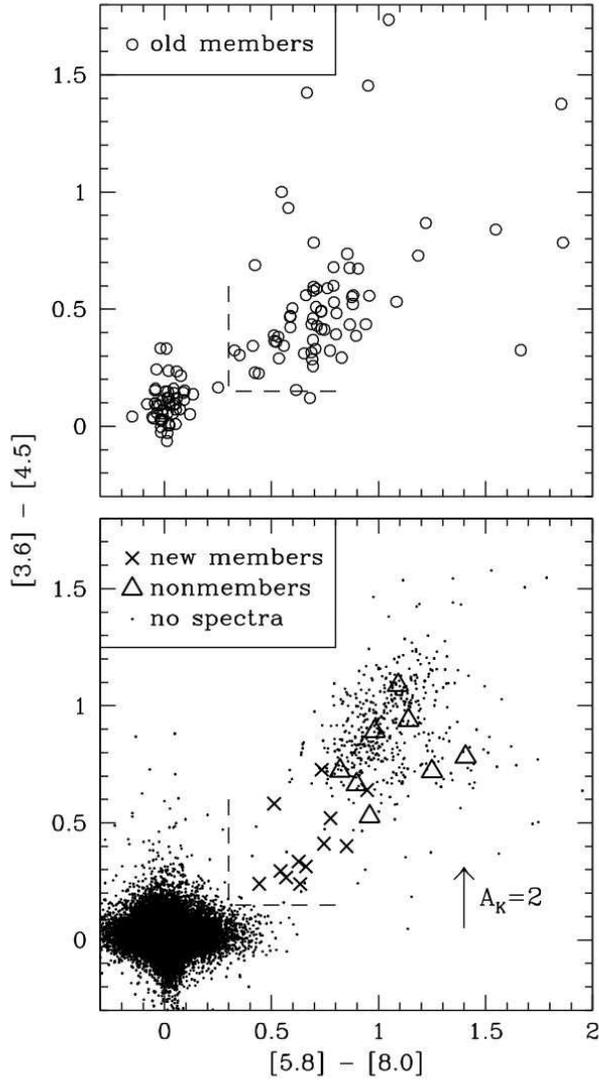}
\caption{
{\it Spitzer} IRAC color-color diagrams for the survey fields in the
Taurus star-forming region indicated in Figure~\ref{fig:map}.
{\it Top}:
Among the previously known members of Taurus
({\it circles}, Table~\ref{tab:old}), the ones with circumstellar disks have 
colors of $[5.8]-[8.0]>0.3$ and $[3.6]-[4.5]>0.15$ ({\it dashed line}).
{\it Bottom}: 
Among the other point sources in the survey,
we obtained spectra of a sample of objects with colors in this diagram
and magnitudes in Figure~\ref{fig:cmd} that are similar to those of the known 
disk-bearing members of Taurus. The candidates classified as
new members ({\it crosses}) and nonmembers ({\it triangles}) are listed in 
Tables~\ref{tab:new} and \ref{tab:non}, respectively. 
Only objects with photometric errors less than 0.1~mag in all four bands
are shown in these two diagrams.
The reddening vector is based on the extinction law from \citet{ind05}.
}
\label{fig:1234}
\end{figure}

To develop criteria for identifying sources in the Archive that could be
new members of Taurus, we examined the colors exhibited by previously 
known members appearing within the IRAC survey.
According to published membership lists for Taurus (\S~\ref{sec:intro}),
149 resolved, previously known members are encompassed by the IRAC images.
One of these members, the protostar IRAS~04368+2557, is not in the Archive
because it is extended and thus was missed by our automated software for
finding point sources. Although the binary IT~Tau is resolved
by the IRAC images, the software identified only one source. 
We measured separate photometry for IT~Tau~A and B by inputting the known 
positions of the two objects into the pipeline and included these data in
the Archive. Among IT~Tau~A and B and the remaining 146 members in the Archive,
118 objects have photometric uncertainties less than 0.1~mag in all four 
IRAC bands. These members are plotted in a diagram of $[3.6]-[4.5]$ versus 
$[5.8]-[8.0]$ in Figure~\ref{fig:1234}.
As demonstrated in the IRAC survey of known Taurus members by \citet{har05}, 
different areas in this diagram correspond to different
classes of mid-IR spectral energy distributions \citep{lw84}.
Stars that lack significant amounts of circumstellar dust (class~III) 
reside near the origin in Figure~\ref{fig:1234} while stars with disks 
(class~II) and stars with both disks and infalling envelopes (class~0 and I)
have progressively redder mid-IR colors. 
The neutral colors of diskless members of Taurus make them indistinguishable
from field stars in IRAC data. However, the other classes are more 
easily separated from most contaminants. In particular, 
\citet{har05} showed that most of the disk-bearing objects in their sample 
exhibited colors of $[3.6]-[4.5]>0.15$ and $[5.8]-[8.0]>0.3$.
Therefore, to identify candidate class~I and II members of Taurus in the new
IRAC survey, we applied these criteria to the $\sim$27,000 sources in the
Archive with photometric errors less than 0.1~mag in all four bands, which are
shown in Figure~\ref{fig:1234}.
We selected for spectroscopy 20 candidates that satisfy these color criteria 
and that appear within the range of magnitudes encompassed by previously
known members ($[3.6]\lesssim13$, Figure~\ref{fig:cmd}).
The star HD~283751 exhibits excess
emission in the IRAC bands, but was classified as a background Be star by
\citet{jh79}, and thus has been excluded from this work.

We also searched for candidate members of Taurus among the IRAC sources that 
have errors less than 0.1~mag in only three bands. The absence of an 
accurate measurement for one of the four bands was typically because of 
saturation (usually at 3.6~\micron), contamination by a cosmic ray, or 
simply low signal-to-noise (usually 5.8 or 8.0~\micron).
To identify candidate class~I and II objects among these IRAC sources, 
we analyzed all possible IRAC color-color diagrams containing three bands in
the same manner as $[3.6]-[4.5]$ versus $[5.8]-[8.0]$.
In this way, we selected one candidate from each of $[3.6]-[4.5]$ versus
$[4.5]-[8.0]$, $[3.6]-[5.8]$ versus $[5.8]-[8.0]$, and 
$[4.5]-[5.8]$ versus $[5.8]-[8.0]$.  The diagram of $[3.6]-[4.5]$ versus 
$[4.5]-[5.8]$ reveals $\sim30$ candidates, but they
are excluded as promising candidates because they are significantly fainter
than expected at 8~\micron\ relative to the other bands for sources with disks.

Spectroscopic observations of the above 23 candidates are described in the next 
section. In \S~\ref{sec:spec}, we classify 13 candidates as Taurus members and
10 candidates as nonmembers.

\section{Spectroscopy of Candidates}
\label{sec:spec}

\subsection{Observations}

We performed optical and near-IR spectroscopy on the 23 candidate Taurus 
members that were selected in the previous section using the 
Marcario Low-Resolution Spectrograph (LRS) on the Hobby-Eberly Telescope (HET),
the Low Dispersion Survey Spectrograph (LDSS-3) on the Magellan~II Telescope,
and SpeX on the Infrared Telescope Facility (IRTF). 
Table~\ref{tab:log} summarizes the observing runs and instrument configurations
for these data. In Tables~\ref{tab:new} and \ref{tab:non}, we indicate the
night on which each object was observed.
The procedures for the collection and reduction of the optical spectra
were similar to those described by \citet{luh04tau}. 
The IR spectra obtained with SpeX \citep{ray03} were reduced with 
the Spextool package \citep{cus04} and corrected for telluric absorption 
with the method from \citet{vac03}.

\subsection{Spectral Classification}
\label{sec:class}

\begin{figure}
\plotone{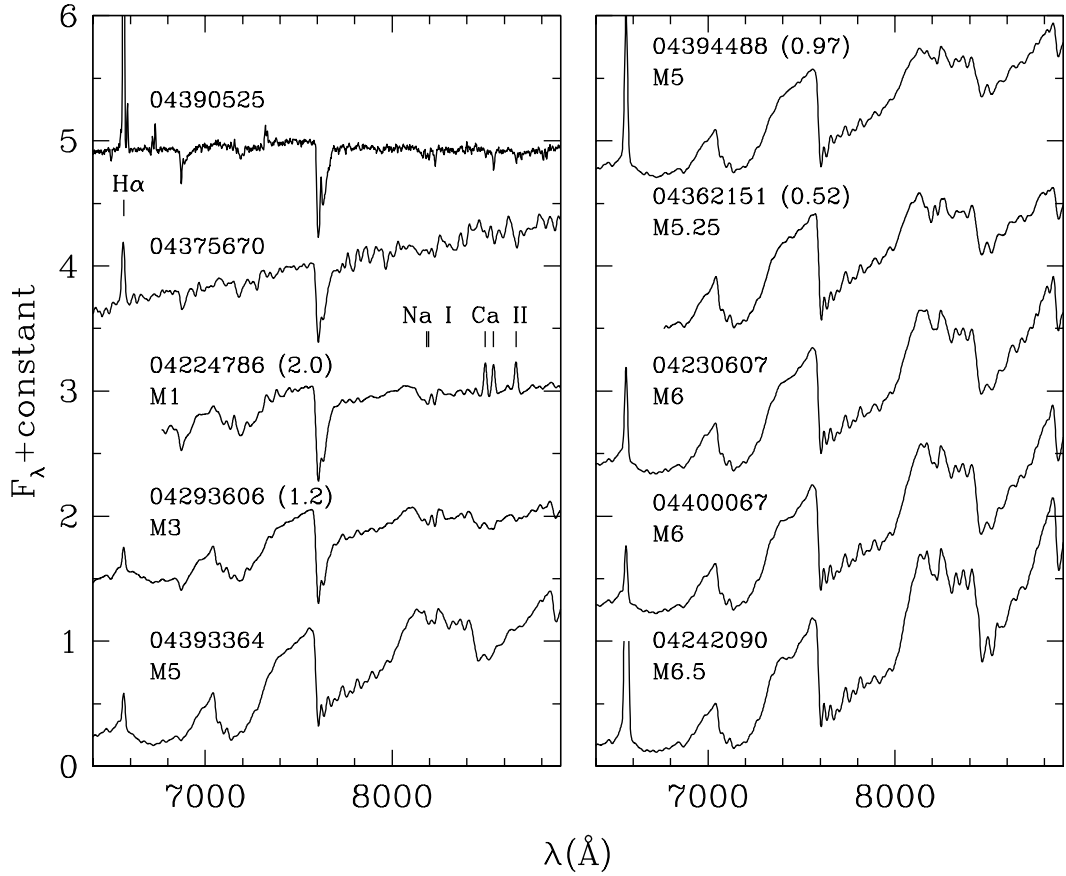}
\caption{
Optical spectra of new members of the Taurus star-forming region discovered
in this work. 
Spectral types could not be measured from the data for the first two stars;
the classifications for the remaining sources are indicated.
The spectra of the M-type objects have been corrected for extinction, which
is quantified in parentheses by the magnitude difference of the reddening
between 0.6 and 0.9~\micron\ ($E(0.6-0.9)$). 
The spectrum of 2MASS~04390525+2337450 is shown at the observed resolution of 
7~\AA\ to facilitate the viewing of its emission lines. 
The remaining data are displayed at a resolution of 18~\AA.
All data are normalized at 7500~\AA. 
}
\label{fig:op}
\end{figure}

\begin{figure}
\plotone{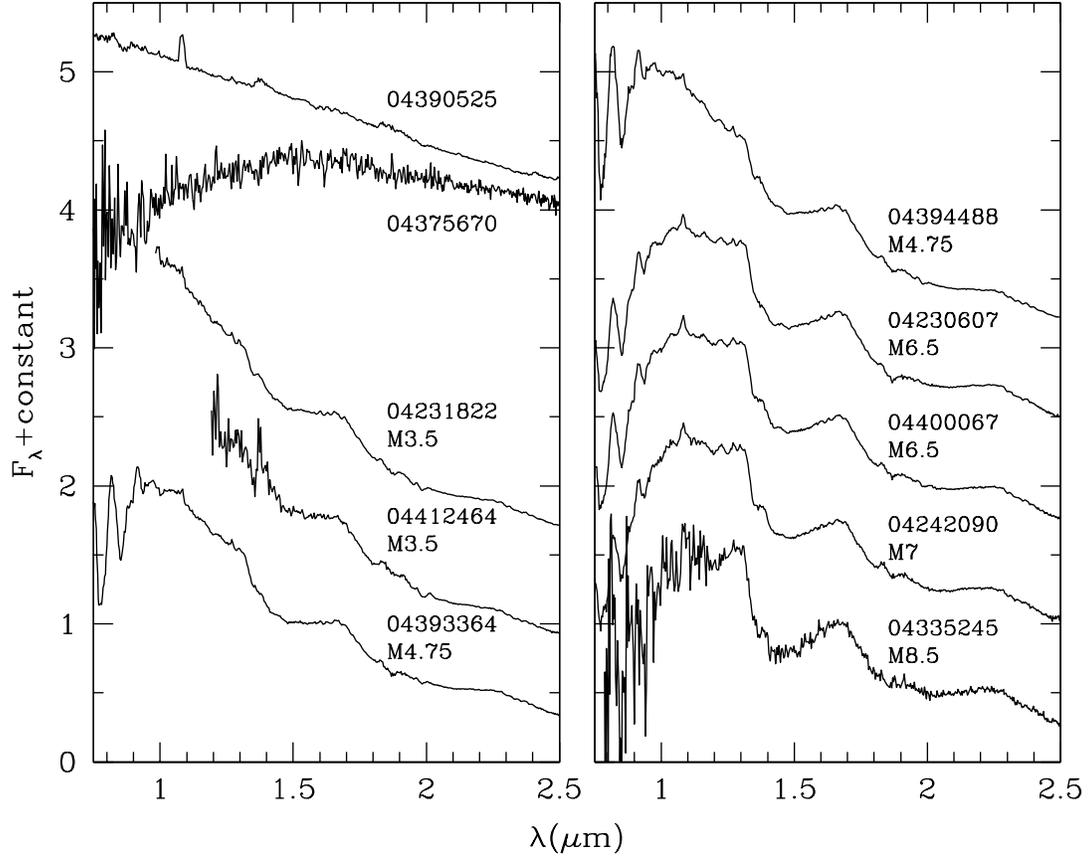}
\caption{
Near-IR spectra of new members of the Taurus star-forming region discovered
in this work. 
Spectral types could not be measured from the data for the first two stars;
the classifications for the remaining sources are indicated.
The spectra for the M-type objects have been dereddened to the same slope 
as measured by the ratios of fluxes at 1.32 and 1.68~\micron.
These data have a resolution of $R=100$ and are normalized at 1.68~\micron.
}
\label{fig:ir}
\end{figure}

To determine if the candidates in our spectroscopic sample are members of
Taurus rather than background sources, particularly red galaxies, and to
measure their spectral types, we applied optical and IR classification 
methods that are identical to those employed in the recent survey for new
Taurus members by \citet{luh06tau}. 
The diagnostics of youth (and thus membership in Taurus) consist of 
emission lines, IR excess emission, gravity-sensitive spectral features,
and reddening. For the new members of Taurus, we measured spectral types 
from the absorption bands of VO and TiO ($\lambda<1.3$~\micron) and 
H$_2$O ($\lambda>1$~\micron) using 
averages of spectra of dwarfs and giants as the spectroscopic
standards at optical wavelengths \citep{luh99} and
optically-classified young objects as the standards at IR wavelengths.
Based on the optical and IR spectra, we classify 13 candidates as members 
of Taurus. Seven of the remaining 10 candidates have emission lines that are 
indicative of active galaxies at redshifts of z=0.07-0.22. 
The final three candidates,
2MASS~J04233697+2526284, J04252936+2654238, and J04385618+2342078, 
lack both detectable stellar features and emission lines from galaxies. 
The spectra of class~I stars can appear featureless because of continuum 
veiling from dust, but these three objects are much bluer than expected for
embedded stars of this kind. It is possible that the presence of edge-on disks
could explain both the blue near-IR colors and red mid-IR colors. 
However, for the purposes of this work, we classify them as nonmembers.
The 2MASS identifications and 2MASS and IRAC photometry for the members and 
nonmembers are provided in Tables~\ref{tab:new} and \ref{tab:non}, respectively.
For each of the new members, we also include our spectral classification 
and the evidence of membership in Taurus. 
The spectra of the new members are presented in order of 
spectral type in Figs.~\ref{fig:op} and \ref{fig:ir}. 
To facilitate the comparison of these spectra, they have been 
corrected for reddening \citep{luh04cha,luh05flam}.
The positions of the new members are indicated in the map of Taurus 
in Figure~\ref{fig:map}.

\subsection{Comments on Individual Sources}

Three of the new members were previously identified as candidate members 
of Taurus through detections of $K$-band excess emission. Those near-IR data 
were collected by \citet{itg96} (ITG) and the astrometry and photometry for the 
candidates, ITG~1, 15, and 40, were reported by \citet{itn99}.
ITG~15 also was identified as a near-IR source by \citet{tam96} and
ITG~40 is probably star 32 from the near-IR survey of \citet{gom94}.
\citet{itt02} reported confirmation of the membership of ITG~15 and 40 
through $K$-band spectroscopy. However, they did not identify any specific 
evidence of youth or membership in their data.
Among the new members, ITG~1 exhibits the reddest IRAC colors and
is probably a class~I object based on its location in Figure~\ref{fig:1234}
\citep{har05}. One of the new members, 2MASS~J04224786+2645530, was
previously detected by IRAS.
Finally, we note that the optical and IR spectra of 2MASS~04390525+2337450 
contain many strong emission lines, including H$\alpha$, He~I, and [O~I], 
[O~II], [N~II], and [S~II], which are indicative of an actively accreting 
classical T~Tauri star with a jet or an outflow.

\section{Survey Completeness}
\label{sec:complete}

\begin{figure}
\epsscale{1}
\plotone{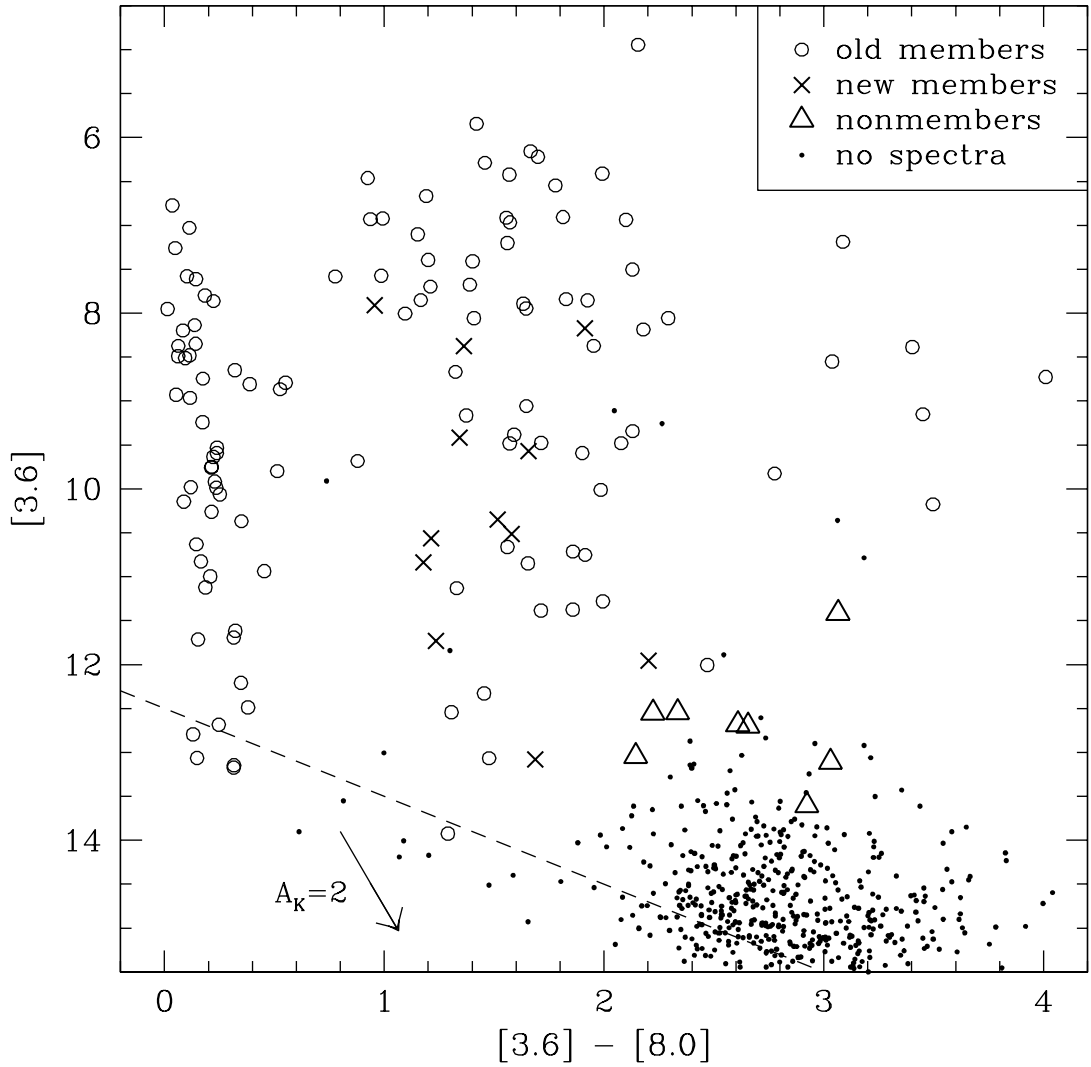}
\caption{
{\it Spitzer} IRAC color-magnitude diagram for the survey fields in the
Taurus star-forming region indicated in Figure~\ref{fig:map}. 
We show the previously known members of Taurus in this survey area 
({\it circles}) and a sample of new sources with colors in 
Figure~\ref{fig:1234} and 
magnitudes in this diagram that are similar to those of known disk-bearing 
members of Taurus, which we have spectroscopically classified as new members
({\it crosses}) and nonmembers ({\it triangles}). We also include the remaining 
sources from Figure~\ref{fig:1234} that have red colors 
($[3.6]-[4.5]>0.15$, $[5.8]-[8.0]>0.3$) and lack spectroscopy ({\it points}).
Only objects with photometric errors less than 0.1~mag in all four bands
are shown in this diagram.
The completeness limit for sources of this kind is shown ({\it dashed line}).
The reddening vector is based on the extinction law from \citet{ind05}.
}
\label{fig:cmd}
\end{figure}

We now evaluate the completeness of both our survey and previous surveys in
Taurus. As mentioned in \S~\ref{sec:cand}, 116 of the 146 previously 
known Taurus members in the IRAC survey fields that were automatically 
identified as point sources have photometric uncertainties less than 0.1~mag
in all four IRAC bands,
and thus satisfy the photometric criteria that we applied to the Archive
in constructing the diagram of $[3.6]-[4.5]$ versus $[5.8]-[8.0]$.
An additional 12 of the known members have accurate photometry in three bands, 
so they satisfy the criteria for the three-band color-color
diagrams with which we selected candidates. 
The remaining 18 known members lack good photometry in three bands
because they are saturated (4), in the field of view
of only two filters (13), or surrounded by extended emission (1). 
If we exclude the members along the edges of the maps with only 2-band coverage
and consider only the area with 4-band coverage, then the color-color diagrams 
employed in selecting candidates contain 128/136 (94\%) of the previously
known members (class~0 through III). This percentage represents an estimate
of the completeness of our photometric sample of class~I and II candidates
for the same range of magnitudes and extinctions exhibited by previously 
known members.  Because we did not obtain spectra of all of these candidates, 
our spectroscopic completeness is lower. However, 
as shown in the diagram of [3.6] versus $[3.6]-[8.0]$ in Figure~\ref{fig:cmd},
the ranges of colors and magnitudes exhibited by previously known Taurus members
contain only $\sim8$ remaining class~I and II candidates that have
not been observed spectroscopically.
Because this survey targeted only class~I and II objects, and roughly 
half of Taurus members fall in these classes, it is likely that another 
10-20 class~III members remain undiscovered in our survey field.

We can quantify the limits in mass and extinction for which our survey is 
complete (for classes~I and II)
by examining the diagram of [3.6] versus $[3.6]-[8.0]$ in Figure~\ref{fig:cmd},
where we show the known members of Taurus and the remaining sources from 
Figure~\ref{fig:1234} that have red colors but lack spectroscopy. 
We include in Figure~\ref{fig:cmd} the photometric completeness limit 
of these data, which was taken to be the value of [8.0] above which the 
number of sources as a function of [8.0] stops rising. 
This limit is computed from [8.0] because it is the least sensitive IRAC band
for the ranges of colors in question.
As demonstrated in Figure~\ref{fig:cmd},
the extinction limit of our survey for class~II members 
($0.5\lesssim[3.6]-[8.0]\lesssim2.2$) is very high, ranging from 
$A_K\sim10$ for 2~$M_\odot$ ($[3.6]\sim6$) 
to $A_K\sim2$ for 0.02~$M_\odot$ ($[3.6]\sim12.5$).
For instance, if the known substellar class~II members of Taurus were reddened
by $A_K=2$, our survey would still detect most of them. 
The completeness also is high for the ranges of colors and magnitudes expected
for class~I objects at stellar masses ($[3.6]-[8.0]\gtrsim2.2$, $[3.6]<11$), 
where there are only two candidates that lack spectroscopy. 
One of these objects, 2MASS~04293209+2430597, is directly within a filament 
of known members and thus is a promising candidate class~I source. 
At the fainter levels expected for class~I brown dwarfs, there remain a large 
number of candidates ($\sim500$) that have not been observed spectroscopically,
most of which are probably galaxies based on their uniform spatial distribution.
In addition to these class~I candidates, Figures~\ref{fig:1234} and
\ref{fig:cmd} contain several objects that lack spectroscopy and have colors
and magnitudes ($0.5\lesssim[3.6]-[8.0]\lesssim1.5$, $[3.6]>13$) 
that are indicative of class~II brown dwarfs at very low masses
($M\sim0.01$~$M_\odot$). 

We can use the new members that we have found to assess the completeness 
of previous surveys.
The recent study by \citet{luh06tau} considered an area of 225~deg$^2$ that
encompasses the entire field imaged by IRAC. The search criteria in that
survey were designed for spectral types later than M6, and two of the new
members in this work are later than this threshold, 2MASS~04242090+2630511 
(M6.5) and 2MASS~04335245+2612548 (M8.5). They were missed by \citet{luh06tau}
because the former is slightly too blue in $I2-K_s$\footnote{$I2$ is the 
second epoch near-IR magnitude in the USNO-B1.0 Catalog \citep{mon03}.} 
and the latter has 2MASS uncertainties that are too large. The fact
that two objects of this kind were not found by \citet{luh06tau} is consistent 
with the level of completeness estimated in that study ($\sim75$\%).
The survey of a smaller area of Taurus by \citet{luh04tau} encompassed three 
of our new members, ITG~1, 15, and 40. 
ITG~40 (M3.5) was not found because it is earlier than the spectral type limit
of $\geq$M4 considered in that search. ITG~15 (M5) is later than this limit,
but it was missed because it is 0.07~mag bluer than the $I-K_s$ threshold
that was adopted for selecting candidates. 
ITG~1 was not identified as a candidate because it was
below the sequence of Taurus members in the color-magnitude diagrams 
from \citet{luh04tau}, which is common for class~I objects because
they are often observed in scattered light. Incompleteness for scattered-light
sources was acknowledged by \citet{luh04tau}.
The IRAC survey also overlaps significantly with the recent optical survey
by \citet{gui06}. Ten of our new members are located within the area 
considered in that study. Four of these objects were not found because
they are earlier than the range of spectral types that was searched ($\geq$M4).
ITG~1 was probably missed in the same manner described for \citet{luh04tau}.
The remaining five new members are within the range of spectral types
and extinctions searched by \citet{gui06}, and consist of 
2MASS~04230607+2801194 (M6), 04242090+2630511 (M6.5), 04335245+2612548 (M8.5), 
04362151+2351165 (M5.25), and ITG~15 (M5).
As discussed by \citet{luh06tau}, the completeness of the survey by
\citet{gui06} is lower than they reported, and these overlooked objects 
further demonstrate this.

\begin{figure}
\plotone{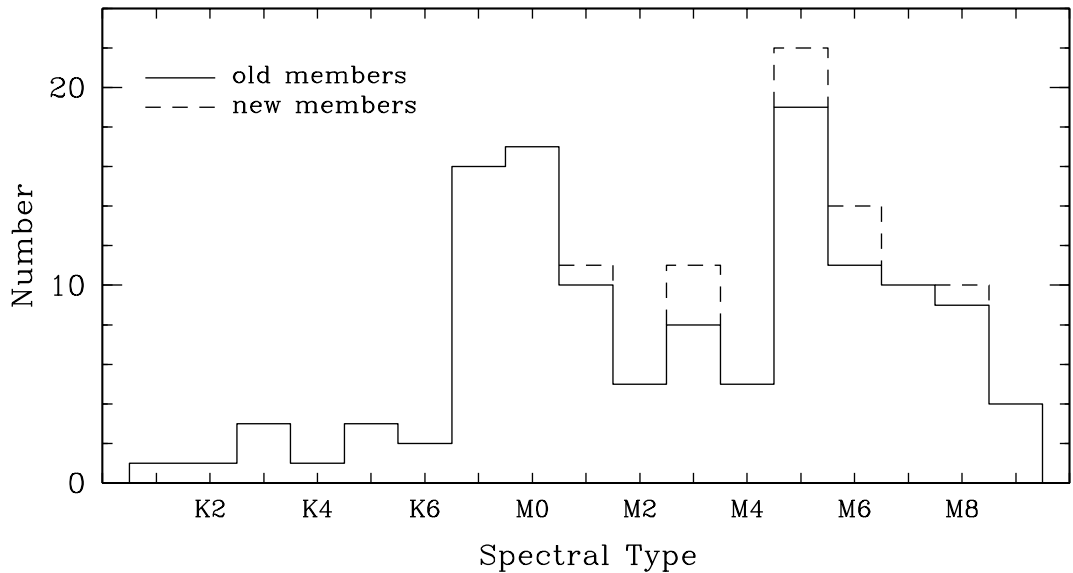}
\caption{
Distribution of spectral types for previously known members of 
the Taurus star-forming region within the survey field in 
Figure~\ref{fig:map} ({\it solid histogram}) and the distribution after 
adding the new members discovered in this work ({\it dashed histogram}).
}
\label{fig:histo}
\end{figure}

The level of completeness of previous membership surveys of Taurus 
at spectral types of M2-M6 has been unknown 
because of a possible gap between the faint 
limits of the wide-field X-ray and objective prism surveys and the
bright limits of deep optical broad-band imaging surveys 
\citep{luh04tau,luh06tau}.
Indeed, 9 of the 11 new members with measured spectral types are 
in this range of spectral types, as illustrated by the distributions
of spectral types in Figure~\ref{fig:histo} for previously known members
within the IRAC survey and for new members.
The color criteria in Figure~\ref{fig:1234} that we used to select
class~I and II candidates are satisfied by 12 new members, 
65 previously known members, and 8 objects that lack spectroscopy
and appear within the ranges of colors and magnitudes of previously known
members, which implies a completeness of $\sim76$-84\% for the 
sum total of all previous surveys across the portion of Taurus imaged by IRAC.

\section{Implications of Survey}
\label{sec:imp}

Because of its sensitivity and large areal coverage, our survey of Taurus
has direct implications for the distributions of Taurus members as a function
of position, mass, and extinction.
As demonstrated in the map of Taurus in Figure~\ref{fig:map}, the spatial
distribution of the new class~I and II sources is similar to that of the 
previously known members in classes~0 through III. 
Among only class~II sources,
most of the new members are clustered near areas of high extinction
in the same manner as the previously known members, while 
three of the new sources in the southeast part of the survey area
are relatively far from dark clouds. 
However, in general, the previous surveys have provided an accurate
measurement of the distribution of class~II objects in Taurus. 
In the portion of our survey that overlaps with areas imaged by \citet{bri02}
and \citet{luh04tau}, we have found only a few new members. As a result,
the measurements of initial mass functions presented in those studies do not
require significant revisions and accurately reflect the Taurus stellar 
population.
The extinction limits of our search for class~II members of Taurus are
very high, and yet the extinctions of the new members are similar to those
of the previously known ones.
Thus, as with position and mass, the previously measured 
distribution of members as a function of extinction also appears to be 
representative of Taurus. In particular, 
some models of brown dwarf disks have suggested that these
disks might have higher scale heights than disks around stars, which would 
result in higher average extinctions for class~II brown dwarfs relative
to class~II stars if the gas and dust are well-mixed \citep{wal04}. 
However, we do not find a population of brown dwarfs that are more highly 
reddened than the previously known ones that were discovered in optical 
and near-IR surveys, which suggests that the disks around brown dwarfs 
do not have higher scale heights, or if they do, that the dust has settled.
Overall, the measured distributions of Taurus members as a function of 
position, mass, and extinction are not changed significantly by our survey.

In addition to identifying new members, the IRAC images also provide mid-IR 
photometry for nearly half of the previously known members of Taurus. 
For several members, a measurement in a given band is unavailable in
the original Archive because a detection was present in only one of 
the two exposures for that band. In these cases, we 
manually inspected the images to determine if the measurement from the one
image was sufficient. In this way, we recovered photometry for LkCa~21, 
J1-665, CFHT~20, FZ~Tau, PSC 04158+2805, and 2MASS~04242090+2630511 in [4.5] and
GO~Tau and IRAS~04239+2436 in [8.0].
In Table~\ref{tab:old}, we present IRAC measurements for 144
known members of Taurus that are within the IRAC fields, excluding 
five class~I members with extended emission. For the latter objects, we 
measured photometry for apertures across a range of radii, which is provided 
in Table~\ref{tab:classI}. These measurements have uncertainties of 
$\sim0.1$~mag. Only one of these class~I sources, IRAS~04248+2612,
has photometry in the Archive that satisfied the criteria 
used in constructing Figure~\ref{fig:1234}. Thus, it is the only one 
appearing in that diagram, and it is plotted with the default photometry
from the Archive rather than the measurements in Table~\ref{tab:classI}.
We note that the colors of the class~I sources in Table~\ref{tab:classI}
vary with aperture size and are distributed all over color-color space, 
including inside the class~II region, which has been predicted 
by models \citep{whi03} and observed in previous IRAC measurements in
Taurus \citep{har05}.

With the IRAC measurements, we can measure the frequency of disks among
stars and brown dwarfs in Taurus. 
For this measurement, we must exclude the new members that we have found
in this work because they were identified through the presence of disk 
emission. These new members cannot be included in a disk fraction measurement 
unless a new survey is performed that encompasses these objects among its 
identified members and that is unbiased in terms of disks.
In addition, we consider only the 118 previously known members that have
photometric uncertainties less than 0.1~mag in all four IRAC bands.
Based on these data, which are shown in Figure~\ref{fig:1234}, 
we measure disk fractions of 
57/93=$61\pm8$\% for the stars ($\leq$M6) and 10/25=$40\pm13$\% for the
brown dwarfs ($>$M6)\footnote{The hydrogen burning mass limit
at ages of 0.5-3~Myr corresponds to a spectral type of $\sim$M6.25
according to the models of \citet{bar98} and \citet{cha00} and the
temperature scale of \citet{luh03ic}.}. 
This measurement for the stars in Taurus is higher than the disk fractions
for stellar members of IC~348 ($33\pm4$\%) and Chamaeleon~I ($45\pm7$\%)
that were measured from IRAC data by \citet{luh05frac}, which is consistent
with the younger age of Taurus \citep[$\tau\sim1$~Myr,][]{bri02,luh03tau} 
relative to these two clusters \citep[$\tau\sim2$~Myr,][]{luh03ic,luh04cha}.
Unlike IC~348 and Chamaeleon~I, the disk fraction of brown dwarfs in
Taurus is lower than that of the stars, but this difference has only marginal
statistical significance and may be the result of the
incompleteness of the current census of Taurus for class~I brown dwarfs.

\section{Conclusions}

We have presented the results of a mid-IR imaging survey of 29.7~deg$^2$ of the 
Taurus star-forming region that was performed with IRAC aboard the 
{\it Spitzer Space Telescope}. 
We used these data to search for new members of Taurus by identifying objects
with colors indicative of circumstellar disks.
Through spectroscopy of 23 of these candidates, we have discovered 13 new
members. One of these objects appears to be a class~I source based on 
its IRAC colors, while two of the new members have spectral types later 
than M6 and thus are likely to be brown dwarfs.
The number of new members found in this survey indicates that the previous
census of Taurus exhibits a completeness of $\sim80$\% for class~I and II 
sources.  
Meanwhile, by considering the number of previously known members of Taurus 
that we have recovered, we estimate that our survey has a completeness 
of $\sim94$\% for class~I and II members.
This completeness extends to high levels of extinction, ranging from
$A_K\sim2$-10 for $M\sim0.02$-2~$M_\odot$.
If brown dwarf disks are more flared than disks around stars, than the
average extinction toward brown dwarfs should be higher than toward stars, 
but we have found no evidence of this in our survey. 
The new members that we have found do not significantly alter the 
distributions of Taurus members as a function of position, mass, 
and extinction that have been measured in previous membership surveys.

In addition to performing a search for new members of Taurus, we have also
measured mid-IR photometry for the 149 previously known members of Taurus 
within the IRAC survey, which comprises nearly half of the known
membership. With these data, we have measured disk fractions of 
$61\pm8$\% and $40\pm13$\% for the stellar ($\leq$M6) 
and substellar ($>$M6) members of Taurus.  The disk fraction for stars in 
Taurus is higher than the fraction for stellar members of 
IC~348 and Chamaeleon~I, which is consistent with the slightly younger age 
of Taurus implied by Hertzsprung-Russell diagrams of these populations.
Meanwhile, the difference in disk fractions for stars and brown dwarfs 
in Taurus has only marginal statistical significance and may be a reflection of 
the incompleteness for class~I brown dwarfs in surveys of this region to date. 
Indeed, $\sim500$ candidate members of Taurus from this IRAC survey
remain unobserved with spectroscopy, most of which would be class~I brown 
dwarfs if confirmed, although a vast majority are probably red galaxies. 

\acknowledgements
K. L. was supported by grant NAG5-11627 from the NASA Long-Term Space 
Astrophysics program.
This work made use of the GLIMPSE data processing pipeline which was supported
by the Spitzer Space Telescope Legacy program through NASA contracts 
1224653 and 1224988.
This work is based in part on observations made with the {\it Spitzer 
Space Telescope}, which is operated by the Jet Propulsion Laboratory, 
California Institute of Technology under a contract with NASA.
This work also is based on observations performed at the IRTF and the HET.
The IRTF is operated by the University of Hawaii under Cooperative 
Agreement no. NCC 5-538 with the National Aeronautics and Space 
Administration, Office of Space Science, Planetary Astronomy Program.
The HET is a joint project of the University of Texas at Austin,
the Pennsylvania State University,  Stanford University,
Ludwig-Maximillians-Universit\"at M\"unchen, and Georg-August-Universit\"at
G\"ottingen.  The HET is named in honor of its principal benefactors,
William P. Hobby and Robert E. Eberly. The Marcario Low-Resolution
Spectrograph at HET is named for Mike Marcario of High Lonesome Optics, who
fabricated several optics for the instrument but died before its completion;
it is a joint project of the Hobby-Eberly Telescope partnership and the
Instituto de Astronom\'{\i}a de la Universidad Nacional Aut\'onoma de M\'exico.
This publication makes use of data products from the Two Micron All
Sky Survey, which is a joint project of the University of Massachusetts
and the Infrared Processing and Analysis Center/California Institute
of Technology, funded by the National Aeronautics and Space
Administration and the National Science Foundation.

\clearpage
\begin{deluxetable}{clllll}
\tablewidth{0pt}
\tablecaption{Observing Log\label{tab:log}}
\tablehead{
\colhead{} &
\colhead{} &
\colhead{} &
\colhead{} &
\colhead{$\lambda$} &
\colhead{} \\
\colhead{Night} &
\colhead{Date} &
\colhead{Telescope + Instrument} &
\colhead{Disperser} &
\colhead{($\mu$m)} &
\colhead{$\lambda/\Delta \lambda$} 
}
\startdata
1 & 2005 Dec 12 & IRTF + SpeX & prism & 0.8-2.5 & 100 \\
2 & 2005 Dec 13 & IRTF + SpeX & prism & 0.8-2.5 & 100 \\
3 & 2005 Dec 14 & IRTF + SpeX & prism & 0.8-2.5 & 100 \\
4 & 2005 Dec 20 & HET + LRS & G3 grism & 0.63-0.91 & 1100 \\
5 & 2005 Dec 23 & HET + LRS & G3 grism & 0.63-0.91 & 1100 \\
6 & 2005 Dec 25 & HET + LRS & G3 grism & 0.63-0.91 & 1100 \\
7 & 2005 Dec 26 & HET + LRS & G3 grism & 0.63-0.91 & 1100 \\
8 & 2006 Feb 9 & Magellan~II + LDSS-3 & VPH red grism & 0.68-1.1 & 1300 \\
9 & 2006 Feb 10 & Magellan~II + LDSS-3 & VPH red grism & 0.58-1 & 1300 \\
\enddata
\end{deluxetable}

\begin{deluxetable}{lllllllllll}
\tabletypesize{\scriptsize}
\tablewidth{0pt}
\tablecaption{New Members of Taurus\label{tab:new}}
\tablehead{
\colhead{} &
\colhead{} &
\colhead{Membership} &
\colhead{} &
\colhead{} &
\colhead{} &
\colhead{} &
\colhead{} &
\colhead{} &
\colhead{} &
\colhead{} \\
\colhead{2MASS\tablenotemark{a}} &
\colhead{Spectral Type\tablenotemark{b}} &
\colhead{Evidence\tablenotemark{c}} &
\colhead{$J-H$\tablenotemark{a}} & 
\colhead{$H-K_s$\tablenotemark{a}} & 
\colhead{$K_s$\tablenotemark{a}} &
\colhead{[3.6]} &
\colhead{[4.5]} &
\colhead{[5.8]} &
\colhead{[8.0]} &
\colhead{Night}}
\startdata
   J04224786+2645530\tablenotemark{d} &         M1 & e,$A_V$,ex &    1.44 &    0.86 &     9.29 &  8.17$\pm$0.06 &  7.45$\pm$0.06 &  6.99$\pm$0.03 &  6.26$\pm$0.02 &   8 \\
   J04230607+2801194 &    M6.5(IR),M6(op) & NaK,e,H$_2$O,ex &    0.63 &    0.41 &    11.20 & 10.56$\pm$0.03 & 10.27$\pm$0.05 &  9.89$\pm$0.04 &  9.35$\pm$0.03 & 1,5 \\
   J04231822+2641156 &       M3.5 &    $A_V$,ex &    1.64 &    0.84 &    10.18 &  9.42$\pm$0.05 &  9.08$\pm$0.04 &  8.70$\pm$0.03 &  8.08$\pm$0.03 &   1 \\
   J04242090+2630511 &    M7(IR),M6.5(op) & e,H$_2$O,ex,NaK &    0.68 &    0.38 &    12.43 & 11.80$\pm$0.05 & 11.42$\pm$0.10 & 10.99$\pm$0.04 & 10.33$\pm$0.03 & 1,4 \\
   J04293606+2435556 &         M3 &    $A_V$,ex &    1.39 &    0.73 &     8.66 &  7.91$\pm$0.04 &  7.67$\pm$0.06 &  7.40$\pm$0.04 &  6.96$\pm$0.03 &   9 \\
   J04335245+2612548 &       M8.5 & $A_V$,H$_2$O,ex &    1.21 &    0.60 &    13.99 & 13.08$\pm$0.05 & 12.56$\pm$0.07 & 12.17$\pm$0.05 & 11.39$\pm$0.04 &   2 \\
   J04362151+2351165 &      M5.25 &   NaK,ex &    0.63 &    0.29 &    12.24 & 11.73$\pm$0.04 & 11.46$\pm$0.05 & 11.06$\pm$0.04 & 10.50$\pm$0.03 &   8 \\
   J04375670+2546229\tablenotemark{e} &          ? &    e,ex &    0.87 &    0.56 &    12.70 & 11.95$\pm$0.06 & 11.31$\pm$0.06 & 10.70$\pm$0.03 &  9.75$\pm$0.03 & 2,5 \\
   J04390525+2337450 &          ? &    e,ex &    1.08 &    0.75 &    11.55 & 10.51$\pm$0.05 & 10.10$\pm$0.04 &  9.68$\pm$0.03 &  8.93$\pm$0.02 & 1,6 \\
   J04393364+2359212 &   M4.75(IR),M5(op) & NaK,H$_2$O,ex &    0.76 &    0.53 &    10.28 &  9.57$\pm$0.03 &  9.17$\pm$0.04 &  8.76$\pm$0.02 &  7.91$\pm$0.03 & 1,7 \\
   J04394488+2601527\tablenotemark{f} &   M4.75(IR),M5(op) & NaK,e,$A_V$,H$_2$O,ex &    1.12 &    0.57 &     8.95 &  8.37$\pm$0.04 &  8.06$\pm$0.05 &  7.67$\pm$0.04 &  7.01$\pm$0.03 & 1,5 \\
   J04400067+2358211 &    M6.5(IR),M6(op) & NaK,e,H$_2$O,ex &    0.56 &    0.39 &    11.48 & 10.84$\pm$0.06 & 10.60$\pm$0.06 & 10.29$\pm$0.04 &  9.66$\pm$0.02 & 1,7 \\
   J04412464+2543530\tablenotemark{g} &       M3.5 &    $A_V$,ex &    3.18 &    1.78 &    11.75 & 10.35$\pm$0.04 &  9.77$\pm$0.04 &  9.35$\pm$0.03 &  8.84$\pm$0.03 &   1 \\
\enddata
\tablenotetext{a}{2MASS Point Source Catalog.}
\tablenotetext{b}{Uncertainties are $\pm0.25$ and $\pm0.5$ subclass for the 
optical and IR types from this work, respectively, unless noted otherwise.}
\tablenotetext{c}{Membership in Taurus is indicated by $A_V\gtrsim1$ and
a position above the main sequence for the distance of Taurus (``$A_V$"),
strong emission lines (``e"), Na~I and K~I strengths intermediate 
between those of dwarfs and giants (``NaK"), IR excess emission (``ex"),
or the shape of the gravity-sensitive steam bands (``H$_2$O").}
\tablenotetext{d}{IRAS 04196+2638.}
\tablenotetext{e}{ITG 1.}
\tablenotetext{f}{ITG 15.}
\tablenotetext{g}{ITG 40.}
\end{deluxetable}

\begin{deluxetable}{lllllllll}
\tabletypesize{\scriptsize}
\tablewidth{0pt}
\tablecaption{Nonmembers\label{tab:non}}
\tablehead{
\colhead{2MASS\tablenotemark{a}} &
\colhead{$J-H$\tablenotemark{a}} & 
\colhead{$H-K_s$\tablenotemark{a}} &
\colhead{$K_s$\tablenotemark{a}} &
\colhead{[3.6]} &
\colhead{[4.5]} &
\colhead{[5.8]} &
\colhead{[8.0]} &
\colhead{Night} 
}
\startdata
   J04233697+2526284 &    1.28 &    0.97 &    14.51 & 13.11$\pm$0.06 & 12.17$\pm$0.07 & 11.22$\pm$0.04 & 10.08$\pm$0.03 &   3 \\
   J04252936+2654238 &    1.23 &    0.22 &    15.85 &   \nodata & 13.20$\pm$0.07 & 12.65$\pm$0.07 & 11.89$\pm$0.05 &   4 \\
   J04253556+2457398 &    0.88 &    0.96 &    14.00 & 12.70$\pm$0.06 & 11.98$\pm$0.05 & 11.29$\pm$0.03 & 10.04$\pm$0.03 &   3 \\
   J04260493+2302201 &    1.06 &    0.95 &    14.61 & 13.61$\pm$0.04 & 12.82$\pm$0.05 & 12.09$\pm$0.04 & 10.68$\pm$0.03 &   2 \\
   J04275447+2424145 &    0.90 &    0.99 &    14.15 & 12.68$\pm$0.07 & 11.79$\pm$0.06 & 11.06$\pm$0.05 & 10.07$\pm$0.02 &   3 \\
   J04340188+2319066 &    1.02 &    0.93 &    14.38 & 12.54$\pm$0.05 & 11.88$\pm$0.06 & 11.11$\pm$0.04 & 10.21$\pm$0.03 &   3 \\
   J04350207+2331415 &    0.93 &    0.55 &    14.21 & 13.05$\pm$0.06 & 12.52$\pm$0.08 & 11.86$\pm$0.04 & 10.90$\pm$0.03 & 2,5 \\
   J04362486+2621473 &    1.36 &    1.13 &    13.89 & 12.55$\pm$0.06 & 11.83$\pm$0.05 & 11.15$\pm$0.04 & 10.33$\pm$0.03 &   2 \\
   J04385618+2342078 &    0.53 &    0.31 &    14.91 & 14.74$\pm$0.06 & 14.53$\pm$0.06 & 14.34$\pm$0.19 & 13.28$\pm$0.07 &   8 \\
   J04443086+2614092 &    1.05 &    1.05 &    13.54 & 11.42$\pm$0.04 & 10.33$\pm$0.04 &  9.44$\pm$0.03 &  8.35$\pm$0.02 & 2,5 \\
\enddata
\tablenotetext{a}{2MASS Point Source Catalog.}
\end{deluxetable}

\begin{deluxetable}{lllllllll}
\tabletypesize{\scriptsize}
\tablewidth{0pt}
\tablecaption{Previously Known Members of Taurus in Spitzer Survey\label{tab:old}}
\tablehead{
\colhead{2MASS\tablenotemark{a}} &
\colhead{Other Name} &
\colhead{$J-H$\tablenotemark{a}} & \colhead{$H-K_s$\tablenotemark{a}}
& \colhead{$K_s$\tablenotemark{a}} &
\colhead{[3.6]} &
\colhead{[4.5]} &
\colhead{[5.8]} &
\colhead{[8.0]} 
}
\startdata
   J04173372+2820468 &      CY Tau  &    0.86 &    0.37 &     8.60 &  7.85$\pm$0.03 &  7.56$\pm$0.06 &  7.22$\pm$0.03 &  6.69$\pm$0.02 \\
   J04174955+2813318 &    KPNO 10 &    0.75 &    0.35 &    10.79 & 10.75$\pm$0.03 & 10.32$\pm$0.03 &  9.78$\pm$0.02 &  8.84$\pm$0.02 \\
   J04174965+2829362 & V410 X-ray 1 &    1.29 &    0.65 &     9.08 &         \nodata &  7.75$\pm$0.05 &         \nodata &  6.43$\pm$0.03 \\
   J04180796+2826036 & V410 X-ray 3 &    0.73 &    0.37 &    10.45 &  9.99$\pm$0.06 &  9.89$\pm$0.05 &  9.77$\pm$0.03 &  9.75$\pm$0.03 \\
   J04181710+2828419 & V410 Anon 13 &    1.29 &    0.70 &    10.96 &         \nodata &  9.82$\pm$0.06 &         \nodata &  8.81$\pm$0.03 \\
   J04182239+2824375 & V410 Anon 24 &    2.89 &    1.53 &    10.73 &  9.80$\pm$0.04 &  9.46$\pm$0.05 &  9.27$\pm$0.03 &  9.28$\pm$0.03 \\
   J04182909+2826191 & V410 Anon 25 &    3.27 &    1.70 &     9.94 &  8.87$\pm$0.04 &  8.53$\pm$0.05 &  8.35$\pm$0.03 &  8.34$\pm$0.03 \\
   J04183030+2743208 &    KPNO 11 &    0.61 &    0.26 &    11.01 & 10.63$\pm$0.04 & 10.53$\pm$0.05 & 10.44$\pm$0.03 & 10.48$\pm$0.04 \\
   J04183110+2827162 &   V410  Tau A+B+C &    0.66 &    0.16 &     7.63 &  7.26$\pm$0.04 &  7.32$\pm$0.04 &  7.22$\pm$0.03 &  7.21$\pm$0.02 \\
   J04183112+2816290 &     DD Tau A+B &    1.15 &    0.80 &     7.88 &  6.41$\pm$0.03 &  5.73$\pm$0.03 &  5.21$\pm$0.03 &  4.42$\pm$0.02 \\
   J04183158+2816585 &     CZ Tau A+B &    0.74 &    0.41 &     9.36 &  8.39$\pm$0.04 &  7.55$\pm$0.05 &  6.53$\pm$0.02 &  4.98$\pm$0.02 \\
   J04183203+2831153 & PSC 04154+2823 &    2.82 &    2.09 &    10.27 &         \nodata &  6.98$\pm$0.04 &         \nodata &  5.47$\pm$0.03 \\
   J04183444+2830302 & V410 X-ray 2 &    3.06 &    1.49 &     9.21 &         \nodata &  8.04$\pm$0.05 &         \nodata &  7.46$\pm$0.03 \\
   J04184023+2824245 & V410 X-ray 4 &    2.68 &    1.28 &     9.69 &  8.81$\pm$0.04 &  8.56$\pm$0.05 &  8.38$\pm$0.03 &  8.42$\pm$0.03 \\
   J04184061+2819155 &    V892 Tau  &    1.73 &    1.23 &     5.79 &  $<$5.04 &         \nodata &  3.58$\pm$0.03 &         \nodata \\
   J04184133+2827250 &        LR1 &    3.30 &    1.87 &    11.05 &  9.48$\pm$0.07 &  8.79$\pm$0.05 &  8.33$\pm$0.03 &  7.91$\pm$0.03 \\
   J04184250+2818498 & V410 X-ray 7 &    1.83 &    0.84 &     9.26 &  8.65$\pm$0.05 &  8.53$\pm$0.05 &  8.39$\pm$0.04 &  8.33$\pm$0.08 \\
   J04184505+2820528 & V410 Anon 20 &    2.99 &    1.48 &    11.93 & 10.94$\pm$0.04 & 10.70$\pm$0.07 & 10.50$\pm$0.04 & 10.48$\pm$0.04 \\
   J04184703+2820073 &    Hubble 4 &    0.92 &    0.34 &     7.29 &  7.03$\pm$0.04 &  7.00$\pm$0.04 &  6.91$\pm$0.03 &  6.92$\pm$0.02 \\
   J04185115+2814332 &     KPNO 2 &    0.68 &    0.49 &    12.75 & 12.21$\pm$0.07 & 12.07$\pm$0.04 & 11.99$\pm$0.05 & 11.86$\pm$0.07 \\
   J04185147+2820264 &  CoKu Tau/1 &    1.38 &    0.52 &    10.97 & 10.28$\pm$0.04 &  8.90$\pm$0.05 &  7.62$\pm$0.02 &  5.77$\pm$0.02 \\
   J04185813+2812234 & PSC 04158+2805 &    1.43 &    1.17 &    11.18 &  9.12$\pm$0.04 &        8.47$\pm$0.09 &  7.76$\pm$0.03 &  6.78$\pm$0.03 \\
   J04190110+2819420 & V410 X-ray 6 &    0.93 &    0.47 &     9.13 &  8.79$\pm$0.05 &  8.63$\pm$0.07 &  8.49$\pm$0.04 &  8.24$\pm$0.02 \\
   J04190126+2802487 &    KPNO 12 &    0.82 &    0.56 &    14.93 & 13.93$\pm$0.06 & 13.54$\pm$0.05 & 13.17$\pm$0.10 & 12.63$\pm$0.06 \\
   J04190197+2822332 & V410 X-ray 5a &    1.20 &    0.63 &    10.15 &  9.59$\pm$0.05 &  9.47$\pm$0.06 &  9.37$\pm$0.03 &  9.36$\pm$0.03 \\
   J04191281+2829330 &     FQ Tau A+B &    0.79 &    0.39 &     9.31 &         \nodata &  8.34$\pm$0.05 &         \nodata &  7.39$\pm$0.02 \\
   J04192625+2826142 &    V819 Tau  &    0.85 &    0.22 &     8.42 &  8.14$\pm$0.03 &  8.13$\pm$0.06 &  8.05$\pm$0.03 &  8.00$\pm$0.03 \\
   J04193545+2827218 &      FR Tau &    0.58 &    0.40 &     9.97 &  9.34$\pm$0.04 &  8.79$\pm$0.06 &  8.17$\pm$0.02 &  7.21$\pm$0.03 \\
   J04194127+2749484 &     LkCa 7A+B &    0.74 &    0.12 &     8.26 &  7.95$\pm$0.04 &  7.98$\pm$0.07 &  7.95$\pm$0.03 &  7.94$\pm$0.03 \\
   J04194148+2716070 & IRAS 04166+2706 &    1.30 &    0.79 &    12.62 & 11.28$\pm$0.06 & 10.35$\pm$0.05 &  9.86$\pm$0.03 &  9.29$\pm$0.03 \\
   J04195844+2709570 & IRAS 04169+2702 &      \nodata &    2.57 &    11.58 &  8.40$\pm$0.05 &         \nodata &  6.28$\pm$0.03 &         \nodata \\
   J04215740+2826355 &      RY Tau  &    1.03 &    0.73 &     5.39 &         \nodata &         \nodata &  3.48$\pm$0.03 &  $<$2.78 \\
   J04215884+2818066 &   HD 283572 &    0.41 &    0.14 &     6.87 &  6.77$\pm$0.04 &  6.80$\pm$0.05 &  6.72$\pm$0.04 &  6.74$\pm$0.03 \\
   J04220313+2825389 &     LkCa 21 &    0.79 &    0.22 &     8.45 &  8.14$\pm$0.03 &        8.15$\pm$0.090  &  8.10$\pm$0.03 &  8.05$\pm$0.03 \\
   J04230776+2805573 & IRAS 04200+2759 &    1.58 &    1.19 &    10.41 &  8.37$\pm$0.04 &  7.77$\pm$0.05 &  7.21$\pm$0.03 &  6.42$\pm$0.02 \\
   J04242646+2649503 &     CFHT 9 &    0.69 &    0.43 &    11.76 & 11.13$\pm$0.04 & 10.85$\pm$0.05 & 10.49$\pm$0.03 &  9.80$\pm$0.03 \\
   J04244457+2610141 & IRAS 04216+2603 &    1.04 &    0.70 &     9.05 &  7.95$\pm$0.07 &  7.49$\pm$0.04 &  6.99$\pm$0.03 &  6.30$\pm$0.03 \\
   J04244506+2701447 &    J1-4423 &    0.63 &    0.26 &    10.46 & 10.14$\pm$0.06 & 10.06$\pm$0.04 & 10.02$\pm$0.04 & 10.06$\pm$0.03 \\
   J04245708+2711565 &      IP Tau  &    0.89 &    0.54 &     8.35 &  7.70$\pm$0.06 &  7.44$\pm$0.05 &  7.18$\pm$0.03 &  6.49$\pm$0.03 \\
   J04251767+2617504 &   J1-4872A &    1.06 &   -0.07 &     8.54 &  8.93$\pm$0.06 &  8.87$\pm$0.07 &  8.84$\pm$0.04 &  8.88$\pm$0.03 \\
   J04251767+2617504 &   J1-4872B &      \nodata &      \nodata &       \nodata &  8.35$\pm$0.04 &  8.32$\pm$0.06 &  8.19$\pm$0.04 &  8.21$\pm$0.03 \\
   J04262939+2624137 &     KPNO 3 &    0.82 &    0.42 &    12.08 & 11.38$\pm$0.05 & 10.90$\pm$0.05 & 10.48$\pm$0.04 &  9.67$\pm$0.02 \\
   J04265352+2606543 &     FV Tau A+B &    1.59 &    0.88 &     7.44 &  6.22$\pm$0.03 &  5.64$\pm$0.04 &  5.22$\pm$0.02 &  4.52$\pm$0.03 \\
   J04265440+2606510 &   FV Tau/c A+B &    1.31 &    0.62 &     8.87 &  7.89$\pm$0.04 &  7.48$\pm$0.04 &  6.99$\pm$0.03 &  6.26$\pm$0.02 \\
   J04265629+2443353 & IRAS 04239+2436 &    3.40 &    2.36 &     9.99 &  7.55$\pm$0.05 &  6.23$\pm$0.04 &  5.32$\pm$0.03 &     4.51$\pm$0.04 \\
   J04265732+2606284 &    KPNO 13 &    1.11 &    0.59 &     9.58 &  8.67$\pm$0.04 &  8.32$\pm$0.06 &  7.90$\pm$0.03 &  7.34$\pm$0.02 \\
   J04270266+2605304 &     DG Tau B &      \nodata &      \nodata &       \nodata &  8.73$\pm$0.06 &  6.99$\pm$0.06 &  5.76$\pm$0.04 &  4.72$\pm$0.03 \\
   J04270280+2542223 &     DF Tau A+B &    0.91 &    0.52 &     6.73 &  $<$5.84 &  $<$5.37 &  5.01$\pm$0.02 &  4.42$\pm$0.03 \\
   J04270469+2606163 &      DG Tau  &    0.97 &    0.73 &     6.99 &  $<$5.73 &         \nodata &  4.46$\pm$0.03 &  3.56$\pm$0.03 \\
   J04272799+2612052 &     KPNO 4 &    0.97 &    0.74 &    13.28 & 12.49$\pm$0.04 & 12.34$\pm$0.06 & 12.20$\pm$0.06 & 12.11$\pm$0.06 \\
   J04274538+2357243 &    CFHT 15 &    0.70 &    0.55 &    13.69 & 13.17$\pm$0.03 & 13.06$\pm$0.07 & 12.95$\pm$0.06 & 12.86$\pm$0.10 \\
   J04284263+2714039 &         \nodata &    1.04 &    0.61 &    10.46 &  9.68$\pm$0.04 &  9.45$\pm$0.06 &  9.23$\pm$0.03 &  8.80$\pm$0.02 \\
   J04290068+2755033 &         \nodata &    0.69 &    0.47 &    12.85 & 12.33$\pm$0.06 & 11.96$\pm$0.05 & 11.57$\pm$0.04 & 10.87$\pm$0.03 \\
   J04290498+2649073 & IRAS 04260+2642 &    1.65 &    1.14 &    11.88 & 10.01$\pm$0.04 &  9.23$\pm$0.07 &  8.72$\pm$0.05 &  8.03$\pm$0.03 \\
   J04292071+2633406 &     J1-507 &    0.73 &    0.30 &     8.79 &  8.51$\pm$0.04 &  8.47$\pm$0.06 &  8.36$\pm$0.03 &  8.42$\pm$0.03 \\
   J04292165+2701259 & IRAS 04263+2654,CFHT 18 &    1.30 &    0.77 &     8.72 &  8.06$\pm$0.04 &  7.63$\pm$0.06 &  7.24$\pm$0.02 &  6.65$\pm$0.02 \\
   J04292373+2433002 &     GV Tau A+B &    1.96 &    1.53 &     8.05 &         \nodata &         \nodata &  $<$2.43 &         \nodata \\
   J04292971+2616532 &     FW Tau A+B &    0.66 &    0.29 &     9.39 &  8.96$\pm$0.04 &  8.91$\pm$0.07 &  8.89$\pm$0.04 &  8.85$\pm$0.03 \\
   J04293008+2439550 & IRAS 04264+2433 &    1.75 &    0.85 &    11.13 & 10.18$\pm$0.09 &  9.39$\pm$0.05 &  8.54$\pm$0.03 &  6.68$\pm$0.02 \\
   J04294155+2632582 &      DH Tau  &    0.94 &    0.65 &     8.18 &  7.58$\pm$0.06 &  7.26$\pm$0.06 &  7.13$\pm$0.03 &  6.80$\pm$0.02 \\
   J04294247+2632493 &     DI Tau A+B &    0.72 &    0.21 &     8.39 &  8.20$\pm$0.04 &  8.20$\pm$0.06 &  8.10$\pm$0.04 &  8.11$\pm$0.03 \\
   J04294568+2630468 &     KPNO 5 &    0.72 &    0.38 &    11.54 & 11.00$\pm$0.06 & 10.90$\pm$0.05 & 10.84$\pm$0.04 & 10.79$\pm$0.04 \\
   J04295156+2606448 &      IQ Tau  &    1.00 &    0.64 &     7.78 &  6.67$\pm$0.06 &  6.31$\pm$0.05 &  6.00$\pm$0.03 &  5.48$\pm$0.03 \\
   J04295950+2433078 &    CFHT 20 &    1.15 &    0.73 &     9.81 &  8.97$\pm$0.05 &  8.60$\pm$0.07 &  8.29$\pm$0.03 &  7.81$\pm$0.03 \\
   J04300724+2608207 &     KPNO 6 &    0.80 &    0.51 &    13.69 & 13.07$\pm$0.06 & 12.74$\pm$0.06 & 12.36$\pm$0.08 & 11.59$\pm$0.07 \\
   J04302365+2359129 &    CFHT 16 &    0.72 &    0.55 &    13.70 & 13.15$\pm$0.05 & 13.10$\pm$0.06 & 12.95$\pm$0.09 & 12.83$\pm$0.09 \\
   J04302961+2426450 &     FX Tau A+B &    0.99 &    0.47 &     7.92 &  6.92$\pm$0.06 &  6.80$\pm$0.06 &  6.61$\pm$0.03 &  5.93$\pm$0.03 \\
   J04304425+2601244 &     DK Tau A+B &    0.96 &    0.66 &     7.10 &  $<$5.76 &  5.67$\pm$0.09 &  5.55$\pm$0.05 &  4.82$\pm$0.03 \\
   J04305028+2300088 & IRAS 04278+2253 &    1.74 &    1.18 &     5.86 &         \nodata &         \nodata &  3.21$\pm$0.03 &         \nodata \\
   J04305137+2442222 &      ZZ Tau  &    0.80 &    0.25 &     8.44 &  8.01$\pm$0.06 &  7.85$\pm$0.06 &  7.53$\pm$0.03 &  6.91$\pm$0.02 \\
   J04305171+2441475 &   ZZ Tau IRS &    1.41 &    1.12 &    10.31 &  8.06$\pm$0.06 &  7.32$\pm$0.06 &  6.62$\pm$0.03 &  5.76$\pm$0.03 \\
   J04305718+2556394 &     KPNO 7 &    0.69 &    0.56 &    13.27 & 12.54$\pm$0.04 & 12.23$\pm$0.03 & 11.89$\pm$0.05 & 11.24$\pm$0.02 \\
   J04311444+2710179 &       JH 56 &    0.67 &    0.24 &     8.79 &  8.74$\pm$0.05 &  8.67$\pm$0.06 &  8.64$\pm$0.03 &  8.57$\pm$0.03 \\
   J04311907+2335047 &         \nodata &    0.79 &    0.52 &    12.20 & 11.69$\pm$0.04 & 11.48$\pm$0.06 & 11.45$\pm$0.05 & 11.38$\pm$0.04 \\
   J04312382+2410529 &   V927 Tau A+B &    0.67 &    0.29 &     8.77 &  8.48$\pm$0.04 &  8.43$\pm$0.06 &  8.36$\pm$0.03 &  8.37$\pm$0.02 \\
   J04312669+2703188 &    CFHT 13 &    0.86 &    0.52 &    13.45 & 12.79$\pm$0.05 & 12.70$\pm$0.04 & 12.58$\pm$0.06 & 12.66$\pm$0.05 \\
   J04315056+2424180 &     HK Tau A+B &    1.20 &    0.66 &     8.59 &  7.57$\pm$0.04 &  7.23$\pm$0.04 &  7.00$\pm$0.03 &  6.59$\pm$0.02 \\
   J04315844+2543299 &     J1-665 &    0.76 &    0.27 &     9.56 &  9.28$\pm$0.03 &  9.24$\pm$0.11 &  9.19$\pm$0.03 &  9.19$\pm$0.03 \\
   J04320329+2528078 &         \nodata &    0.61 &    0.39 &    10.72 & 10.26$\pm$0.06 & 10.16$\pm$0.05 & 10.07$\pm$0.04 & 10.05$\pm$0.03 \\
   J04321540+2428597 &   Haro 6-13 &    1.92 &    1.22 &     8.10 &  6.29$\pm$0.05 &  5.78$\pm$0.05 &  5.43$\pm$0.03 &  4.83$\pm$0.03 \\
   J04321786+2422149 &     CFHT 7 &    0.75 &    0.41 &    10.38 &  9.92$\pm$0.04 &  9.77$\pm$0.05 &  9.70$\pm$0.04 &  9.69$\pm$0.03 \\
   J04321885+2422271 &   V928 Tau A+B &    1.11 &    0.33 &     8.11 &  7.80$\pm$0.04 &  7.79$\pm$0.04 &  7.63$\pm$0.03 &  7.61$\pm$0.03 \\
   J04322329+2403013 &         \nodata &    0.64 &    0.36 &    11.33 & 10.83$\pm$0.06 & 10.77$\pm$0.05 & 10.67$\pm$0.04 & 10.66$\pm$0.03 \\
   J04323058+2419572 &      FY Tau  &    1.31 &    0.62 &     8.05 &  7.10$\pm$0.03 &  6.71$\pm$0.04 &  6.46$\pm$0.03 &  5.95$\pm$0.03 \\
   J04323176+2420029 &      FZ Tau  &    1.49 &    1.05 &     7.35 &  6.22$\pm$0.04 &        5.69$\pm$0.08 &  5.21$\pm$0.03 &  4.55$\pm$0.03 \\
   J04323205+2257266 & IRAS 04295+2251 &    2.91 &    1.84 &    10.14 &  8.55$\pm$0.04 &  7.68$\pm$0.06 &  6.73$\pm$0.03 &  5.51$\pm$0.03 \\
   J04324282+2552314 &    UZ Tau Ba+Bb &    1.41 &    0.53 &     7.47 &  7.67$\pm$0.05 &  7.34$\pm$0.05 &  6.99$\pm$0.04 &  6.29$\pm$0.03 \\
   J04324303+2552311 &     UZ Tau A &      \nodata &      \nodata &     7.35 &  6.55$\pm$0.04 &  6.02$\pm$0.05 &  5.56$\pm$0.03 &  4.77$\pm$0.02 \\
   J04324911+2253027 &      JH 112 &    1.24 &    0.83 &     8.17 &  7.33$\pm$0.07 &         \nodata &  6.76$\pm$0.03 &         \nodata \\
   J04325026+2422115 &     CFHT 5 &    1.74 &    0.94 &    11.28 & 10.37$\pm$0.04 & 10.21$\pm$0.05 & 10.06$\pm$0.04 & 10.02$\pm$0.03 \\
   J04330197+2421000 &       MHO 8 &    0.72 &    0.41 &     9.73 &  9.24$\pm$0.05 &  9.13$\pm$0.04 &  9.06$\pm$0.03 &  9.07$\pm$0.03 \\
   J04330622+2409339 &     GH Tau A+B &    0.88 &    0.44 &     7.79 &  6.93$\pm$0.07 &  6.70$\pm$0.04 &  6.43$\pm$0.03 &  5.99$\pm$0.02 \\
   J04330664+2409549 &   V807 Tau A+B &    0.79 &    0.40 &     6.96 &  6.46$\pm$0.04 &  6.16$\pm$0.05 &  5.89$\pm$0.03 &  5.54$\pm$0.03 \\
   J04330781+2616066 &    KPNO 14 &    1.10 &    0.54 &    10.27 &  9.75$\pm$0.05 &  9.60$\pm$0.05 &  9.49$\pm$0.04 &  9.53$\pm$0.03 \\
   J04331003+2433433 &    V830 Tau  &    0.71 &    0.19 &     8.42 &  8.37$\pm$0.04 &  8.34$\pm$0.06 &  8.29$\pm$0.03 &  8.31$\pm$0.03 \\
   J04331435+2614235 & IRAS 04301+2608 &    1.41 &    0.73 &    12.50 & 12.01$\pm$0.04 & 11.68$\pm$0.06 & 11.20$\pm$0.04 &  9.54$\pm$0.03 \\
   J04333405+2421170 &      GI Tau  &    0.92 &    0.53 &     7.89 &  6.94$\pm$0.08 &  6.26$\pm$0.06 &  5.74$\pm$0.04 &  4.84$\pm$0.03 \\
   J04333456+2421058 &      GK Tau  &    0.95 &    0.64 &     7.47 &  6.42$\pm$0.05 &  6.04$\pm$0.05 &  5.75$\pm$0.03 &  4.85$\pm$0.03 \\
   J04333678+2609492 &     IS Tau A+B &    1.03 &    0.65 &     8.64 &  7.84$\pm$0.04 &  7.41$\pm$0.06 &  6.88$\pm$0.04 &  6.01$\pm$0.02 \\
   J04333906+2520382 &      DL Tau  &    0.95 &    0.72 &     7.96 &  6.91$\pm$0.04 &  6.32$\pm$0.04 &  5.85$\pm$0.03 &  5.09$\pm$0.03 \\
   J04334291+2526470 &         \nodata &    0.79 &    0.52 &    13.33 & 12.68$\pm$0.06 & 12.54$\pm$0.06 & 12.53$\pm$0.06 & 12.44$\pm$0.08 \\
   J04335470+2613275 &     IT Tau A &    1.28 &    0.73 &     7.86 &  7.39$\pm$0.04 &  7.03$\pm$0.05 &  6.71$\pm$0.04 &  6.19$\pm$0.02 \\
   \nodata &     IT Tau B &    \nodata & \nodata &     \nodata &  9.06$\pm$0.07 &  8.57$\pm$0.09 &  8.14$\pm$0.06 &  7.41$\pm$0.04 \\
   J04345542+2428531 &      AA Tau  &    0.89 &    0.50 &     8.05 &  7.20$\pm$0.04 &  6.77$\pm$0.04 &  6.35$\pm$0.04 &  5.64$\pm$0.03 \\
   J04350850+2311398 &    CFHT 11 &    0.59 &    0.35 &    11.59 & 11.12$\pm$0.07 & 11.03$\pm$0.06 & 10.98$\pm$0.04 & 10.94$\pm$0.04 \\
   J04352737+2414589 &      DN Tau  &    0.80 &    0.33 &     8.02 &  7.41$\pm$0.05 &  7.09$\pm$0.06 &  6.69$\pm$0.04 &  6.01$\pm$0.03 \\
   J04354093+2411087 &  CoKu Tau 3A+B &    1.53 &    0.79 &     8.41 &  6.91$\pm$0.09 &  6.63$\pm$0.11 &  6.41$\pm$0.05 &  5.64$\pm$0.03 \\
   J04354526+2737130 &         \nodata &    0.77 &    0.53 &    13.71 & 13.06$\pm$0.04 & 13.02$\pm$0.06 & 12.77$\pm$0.07 & 12.92$\pm$0.09 \\
   J04355109+2252401 &    KPNO 15 &    0.96 &    0.34 &    10.01 &  9.74$\pm$0.04 &         \nodata &  9.59$\pm$0.03 &         \nodata \\
   J04355277+2254231 &      HP Tau  &    1.08 &    0.84 &     7.62 &         \nodata &         \nodata &  5.53$\pm$0.02 &         \nodata \\
   J04355349+2254089 &   HP Tau/G3 &    0.89 &    0.36 &     8.80 &  8.53$\pm$0.04 &         \nodata &  8.41$\pm$0.03 &         \nodata \\
   J04355415+2254134 &   HP Tau/G2 &    0.61 &    0.25 &     7.23 &  7.05$\pm$0.04 &         \nodata &  6.99$\pm$0.03 &         \nodata \\
   J04355684+2254360 &  Haro 6-28 A+B &    1.09 &    0.52 &     9.53 &  8.55$\pm$0.04 &         \nodata &  7.87$\pm$0.02 &         \nodata \\
   J04361038+2259560 &     CFHT 2 &    0.99 &    0.59 &    12.17 & 11.62$\pm$0.05 & 11.38$\pm$0.05 & 11.35$\pm$0.04 & 11.29$\pm$0.03 \\
   J04361909+2542589 &     LkCa 14 &    0.62 &    0.13 &     8.58 &  8.49$\pm$0.05 &  8.49$\pm$0.06 &  8.45$\pm$0.02 &  8.43$\pm$0.02 \\
   J04363893+2258119 &     CFHT 3 &    0.86 &    0.49 &    12.37 & 11.71$\pm$0.05 & 11.62$\pm$0.05 & 11.54$\pm$0.04 & 11.56$\pm$0.04 \\
   J04380083+2558572 &       ITG 2 &    0.92 &    0.53 &    10.10 &  9.53$\pm$0.04 &  9.41$\pm$0.04 &  9.31$\pm$0.03 &  9.29$\pm$0.03 \\
   J04381486+2611399 &         \nodata &    1.05 &    1.15 &    12.98 & 10.71$\pm$0.03 & 10.12$\pm$0.05 &  9.55$\pm$0.03 &  8.86$\pm$0.03 \\
   J04382134+2609137 &      GM Tau  &    1.22 &    0.95 &    10.63 &  9.16$\pm$0.05 &  8.70$\pm$0.07 &  8.38$\pm$0.03 &  7.79$\pm$0.02 \\
   J04382858+2610494 &      DO Tau  &    1.23 &    0.94 &     7.30 &  6.16$\pm$0.03 &  5.65$\pm$0.05 &  5.20$\pm$0.03 &  4.49$\pm$0.03 \\
   J04383528+2610386 &     HV Tau A+B &    0.94 &    0.38 &     7.91 &  7.61$\pm$0.04 &  7.54$\pm$0.05 &  7.46$\pm$0.03 &  7.47$\pm$0.03 \\
   J04390396+2544264 &     CFHT 6 &    0.81 &    0.47 &    11.37 & 10.66$\pm$0.05 & 10.37$\pm$0.06 &  9.93$\pm$0.03 &  9.10$\pm$0.03 \\
   J04392090+2545021 &     GN Tau A+B &    1.30 &    0.83 &     8.06 &  6.97$\pm$0.04 &  6.48$\pm$0.05 &  6.13$\pm$0.03 &  5.39$\pm$0.03 \\
   J04393519+2541447 & IRAS 04365+2535 &      \nodata &    2.92 &    10.84 &  7.19$\pm$0.06 &  5.76$\pm$0.04 &  4.77$\pm$0.03 &  4.10$\pm$0.03 \\
   J04394748+2601407 &     CFHT 4 &    1.16 &    0.68 &    10.33 &  9.38$\pm$0.05 &  8.97$\pm$0.06 &  8.54$\pm$0.02 &  7.79$\pm$0.03 \\
   J04395574+2545020 &   IC2087IR &    2.62 &    1.78 &     6.28 &  $<$4.95 &  $<$3.94 &  3.34$\pm$0.03 &  $<$2.79 \\
   J04400174+2556292 &    CFHT 17 &    1.58 &    0.88 &    10.76 & 10.06$\pm$0.05 &  9.90$\pm$0.05 &  9.77$\pm$0.03 &  9.81$\pm$0.03 \\
   J04400800+2605253 & IRAS 04370+2559 &    2.16 &    1.38 &     8.87 &  7.85$\pm$0.04 &  7.30$\pm$0.06 &  6.80$\pm$0.03 &  5.93$\pm$0.03 \\
   J04403979+2519061 &         \nodata &    1.03 &    0.55 &    10.24 &  9.76$\pm$0.05 &  9.61$\pm$0.05 &  9.55$\pm$0.03 &  9.54$\pm$0.03 \\
   J04404950+2551191 &      JH 223 &    0.83 &    0.43 &     9.49 &  8.86$\pm$0.11 &  8.49$\pm$0.07 &  8.17$\pm$0.05 &  7.67$\pm$0.02 \\
   J04410424+2557561 &   Haro 6-32 &    0.68 &    0.31 &     9.95 &  9.64$\pm$0.05 &  9.49$\pm$0.06 &  9.46$\pm$0.02 &  9.41$\pm$0.03 \\
   J04410826+2556074 &     ITG 33A &    1.59 &    1.06 &    11.09 &  9.59$\pm$0.04 &  9.01$\pm$0.06 &  8.40$\pm$0.02 &  7.69$\pm$0.03 \\
   J04411078+2555116 & ITG 34,CFHT 8 &    1.07 &    0.67 &    11.45 & 10.85$\pm$0.06 & 10.29$\pm$0.03 &  9.86$\pm$0.04 &  9.19$\pm$0.02 \\
   J04411267+2546354 & IRAS 04381+2540 &    3.01 &    2.60 &    11.54 &  9.15$\pm$0.04 &  7.70$\pm$0.06 &  6.65$\pm$0.02 &  5.70$\pm$0.03 \\
   J04413882+2556267 & IRAS 04385+2550 &    1.73 &    0.92 &     9.20 &  8.18$\pm$0.04 &  7.65$\pm$0.04 &  7.09$\pm$0.03 &  6.00$\pm$0.03 \\
   J04414825+2534304 &         \nodata &    0.93 &    0.58 &    12.22 & 11.38$\pm$0.05 & 10.85$\pm$0.04 & 10.40$\pm$0.04 &  9.52$\pm$0.03 \\
   J04420548+2522562 & LkHa 332/G2A+B &    1.12 &    0.44 &     8.23 &  7.86$\pm$0.05 &  7.79$\pm$0.06 &  7.69$\pm$0.03 &  7.64$\pm$0.02 \\
   J04420732+2523032 & LkHa 332/G1A+B &    1.18 &    0.46 &     7.95 &  7.58$\pm$0.03 &  7.54$\pm$0.05 &  7.42$\pm$0.04 &  7.48$\pm$0.04 \\
   J04420777+2523118 &   V955 Tau A+B &    1.21 &    0.66 &     7.94 &  6.91$\pm$0.07 &  6.48$\pm$0.05 &  6.04$\pm$0.03 &  5.36$\pm$0.03 \\
   J04422101+2520343 &     CIDA 7 &    0.82 &    0.41 &    10.17 &  9.48$\pm$0.04 &  9.08$\pm$0.05 &  8.56$\pm$0.03 &  7.76$\pm$0.02 \\
   J04423769+2515374 &      DP Tau  &    1.31 &    0.93 &     8.76 &  7.50$\pm$0.03 &  6.83$\pm$0.05 &  6.24$\pm$0.04 &  5.37$\pm$0.03 \\
   J04430309+2520187 &      GO Tau  &    0.94 &    0.44 &     9.33 &  8.88$\pm$0.05 &  8.59$\pm$0.07 &  8.14$\pm$0.04 &  7.41$\pm$0.04 \\
   J04442713+2512164 & IRAS 04414+2506 &    0.84 &    0.60 &    10.76 &  9.48$\pm$0.05 &  8.92$\pm$0.06 &  8.28$\pm$0.03 &  7.40$\pm$0.03 \\
   J04464260+2459034 & RXJ 04467+2459 &    0.59 &    0.33 &    10.34 &  9.98$\pm$0.05 &  9.88$\pm$0.04 &  9.82$\pm$0.03 &  9.86$\pm$0.02 \\
\enddata
\tablenotetext{a}{2MASS Point Source Catalog.}
\end{deluxetable}

\begin{deluxetable}{lllllll}
\tablewidth{0pt}
\tablecaption{Members of Taurus with Extended Emission\label{tab:classI}}
\tablehead{
\colhead{2MASS\tablenotemark{a}} &
\colhead{IRAS} &
\colhead{Radius ($\arcsec$)} &
\colhead{[3.6]} &
\colhead{[4.5]} &
\colhead{[5.8]} &
\colhead{[8.0]} 
}
\startdata
04275730+2619183 & 04248+2612 &   4 &  9.71 &  9.04 &  8.33 &  7.25 \\
 & & 6 &  9.51 &  8.89 &  8.22 &  7.10 \\
 & & 13 &  9.29 &  8.72 &  8.09 &  7.01 \\
 & & 26 &  9.09 &  8.57 &  7.92 &  6.92 \\
04331650+2253204 & 04302+2247 & 4 &  9.97 &  \nodata &  9.57 & \nodata \\
 & &  6 &  9.80   &   \nodata &    9.41   &     \nodata \\
 & & 13 &   9.63  &   \nodata &    9.26    &    \nodata \\
 & & 26 &   9.49  &   \nodata &    9.09    &    \nodata \\
04353539+2408194 & 04325+2402 & 4 &  9.80 &  9.15 &  8.98 &  8.59 \\
 & & 6 &  9.54  & 8.92  & 8.77  & 8.37 \\
 & & 13 &  9.16 &  8.60 &  8.45 &  8.15 \\
 & & 26 &  8.94 &  8.42 &  8.23 &  8.05 \\
04391389+2553208 & 04361+2547  &  4 &  8.04 &  7.10 & 6.51 &  4.97  \\
 & & 6 &  7.95 &  7.01  & 6.43  & 4.85 \\
 & & 13 &  7.80 &  6.88 &  6.34 &  4.76 \\
 & & 26 &  7.66 &  6.76 &  6.21 &  4.70 \\
 & & 35 &  7.57 &  6.77 &  6.11 &  4.66 \\
\nodata & 04368+2557  &  4 & 12.70 & 10.74 &  9.74 &  9.47 \\
 & & 6 & 11.89 & 10.16  & 9.25 & 9.02 \\
 & & 13 & 10.48 &  9.13 &  8.40 &  8.33 \\
 & & 26 &  9.32 &  8.22 &  7.59 &  7.68 \\
 & & 60 &  8.33 &  7.44 &  6.91 &  7.41 \\
 & & 100 &  8.07 &  7.28 &  6.60 &  7.23 \\
\enddata
\tablenotetext{a}{2MASS Point Source Catalog.}
\end{deluxetable}

\clearpage

\end{document}